\documentstyle[11pt]{article}

\def\slashchar#1{\setbox0=\hbox{$#1$}           
   \dimen0=\wd0                                 
   \setbox1=\hbox{/} \dimen1=\wd1               
   \ifdim\dimen0>\dimen1                        
      \rlap{\hbox to \dimen0{\hfil/\hfil}}      
      #1                                        
   \else                                        
      \rlap{\hbox to \dimen1{\hfil$#1$\hfil}}   
      /                                         
   \fi}
\begin{document}
\begin{titlepage}
\begin{center}
April, 1997      \hfill     BUHEP-97-14\\
\vskip 0.2 in
{\large \bf Reparameterization Invariance to all Orders\\
 in Heavy Quark   Effective Theory}

\vskip .2 in
         {\bf   Raman Sundrum\footnote{email: sundrum@budoe.bu.edu.}}
        \vskip 0.3 cm
       {\it Department of Physics \\
Boston University \\
Boston, MA 02215, USA}
 \vskip 0.7 cm 

\begin{abstract}
Heavy Quark Effective Theory  splits a heavy quark momentum 
into a large fixed momentum and a variable residual momentum, 
$p_{\mu} = m_Q v_{\mu} + k_{\mu}$. It thereby suffers a redundancy 
of description corresponding to small changes in the choice of the 
fixed velocity, $v_{\mu}$. The fact that full QCD is manifestly 
$v$-independent should lead to a non-trivial constraint on the form 
of the effective theory, known as Reparameterization Invariance. For 
spin-$1/2$ quarks, the precise form of the constraint and its solution
at the level of the effective lagrangian have proven to be rather 
subtle, and the original proposal by Luke and Manohar has been 
questioned. In this paper I employ a version of Heavy Quark Effective 
Theory containing the ``anti-particle'' field as a non-propagating 
auxiliary field, which greatly simplifies keeping track of 
$v$-dependence. This permits a very simple derivation of 
Reparameterization Invariance from first principles. 
 The auxiliary field can also be integrated out to return to 
the standard formulation of the effective theory, but with the
effective lagrangian now satisfying the full reparameterization 
constraint. I compare this result with earlier proposals. 
\end{abstract}
\end{center}

\end{titlepage}

\section{Introduction}

Heavy Quark Effective Theory (HQET) provides a systematic expansion of
QCD amplitudes in powers of (momentum transfer)$/m$, where $m$ is a
heavy quark mass. 
Calculations are very conveniently done in terms of an
effective lagrangian which is also organized in powers of $1/m$. 
 Ref. \cite{grinstein} contains a review and references
to  the original literature.
There are many interesting processes where the momentum
transfers between a heavy quark and the light quarks and gluons are 
of order $\Lambda_{QCD}$. For such situations and for $m \gg \Lambda_{QCD}$, 
the heavy quark {\it velocity}
is approximately constant. This is formally reflected in setting up the 
$1/m$-expansion  by splitting the heavy quark momentum into two pieces,
\begin{equation}
p_{\mu} = mv_{\mu} + k_{\mu},
\end{equation}
where $v_{\mu}$ is a {\it fixed} four-velocity ($v^2 = 1$), and
$k_{\mu}$ is a variable residual momentum of order
$\Lambda_{QCD}$. While this split is very useful for book-keeping,
clearly there is an arbitrariness of order $\Lambda_{QCD}/m$ in the
precise choice of $v_{\mu}$ for any process \cite{dugan}.
 Formally we need just
note the redundancy of the momentum decomposition under {\it infinitesmal}
shifts in the fixed velocity, 
\begin{equation}
v' = v + \delta v,  ~~~~~ ~v.\delta v = 0.
\end{equation}
QCD is manifestly invariant under such infinitesimal reparameterizations,
so a systematic approximation to full QCD such as HQET must also 
reflect this fact. This constraint on HQET is known as
Reparameterization Invariance (RPI). Luke and Manohar \cite{luke} proposed that
RPI should directly constrain the form of the lagrangian, like any other
symmetry. Since the HQET
 lagrangian must be determined order by order in $\alpha_s$ and $1/m$ by
 matching to full QCD, there is a significant reduction of work and
 increase in understanding if one 
 can first enforce RPI on the general form of the effective lagrangian. 

The precise form of the RPI constraint on the HQET lagrangian for
spin-$1/2$ quarks has been rather difficult to identify and prove. The
original proposal by Luke and Manohar \cite{luke} 
for the form of the RPI-constrained 
lagrangian was shown by Chen \cite{chen} 
 to be inconsistent with the direct matching
computation already at tree-level. It is possible that the proposal is
valid up to field redefinitions (see ref. \cite{manohar} for an
application of RPI by this means) but this is less useful.
 Chen proposed another form for the RPI
constraint consistent with tree-level matching, but its general
validity has been unclear because the derivation effectively depends
on the use of  the classical
equations of motion. Kilian and Ohl \cite{kilian} obtained a RPI form for
the HQET lagrangian valid to all orders, and 
which reduces at tree-level to Chen's version. 
I believe this to be 
correct and part of my present purpose is to give a
simpler derivation and better understanding of this result.
RPI in HQET has most recently been
investigated by Finkemeier, Georgi and McIrvin \cite{finkemeier},
 who showed that the constraints from Chen's
version of RPI are correct to order $1/m^2$. The framework of the
present paper will clarify why this happens even though Chen's RPI is
wrong in general (see section 6). 
Ref. \cite{finkemeier} also showed that there is a field 
redefinition  (which does not affect the S-matrix) taking Chen's constraints
 to those predicted by Luke and Manohar's RPI, to
 order $1/m^2$. 

In order to understand the nature of the difficulty and how to resolve
it, let us briefly review the essentials of the HQET procedure. {\it A
  priori} the correct procedure for calculating an amplitude to some
order in $1/m$ (and in $\alpha_s$) is to compute the
relevant Feynman diagrams of full QCD and to expand the result in powers
of $1/m$. HQET works essentially in the reverse order, which is the key
to its conceptual and computational advantages. Having
separated out the fixed large momentum as in eq. (1), the QCD vertices
and propagators are {\it first} expanded in powers of $1/m$ and then
used to compute Feynman diagrams. Because of the different ultraviolet
divergence structures that arise depending on whether one expands in
$1/m$ first or last, this naive version of HQET only agrees at
tree-level with full QCD. However, the two procedures can be made to
agree (``match'') by adding local terms to the HQET
lagrangian. Once the extra terms in ${\cal L}_{HQET}$ are determined (to
any particular order) by matching to some number of full QCD amplitudes,
they are universal, and can be used to compute other amplitudes without
further reference to full QCD. This procedure is quite general and
applies even if the quarks were spinless. In the spinless case
there is no difficulty in tracking $v$-dependence, deriving the form of
RPI for the effective theory, and enforcing the constraint on the form
of the effective lagrangian \cite{luke}.

For spin-$1/2$ quarks there is an extra complication. 
When the quark propagator is expanded for
$k_{\mu} \ll m$ in eq. (1), one finds
\begin{equation}
\frac{1}{\slashchar{p} - m + i \epsilon} = 
\frac{1 + \slashchar{v}}{2} ~\frac{1}{k.v  + i \epsilon} + {\cal
  O}(\frac{1}{m}).
\end{equation}
This shows that in the heavy quark limit $m \rightarrow \infty$, not all
four spinor degrees of freedom propagate, but only the two components
projected out by $(1 + \slashchar{v})/2$. The usual practice is to
interpolate these spinor modes in HQET with a heavy quark field
$\psi_{+v}$, 
\begin{equation}
\slashchar{v} \psi_{+v} = \psi_{+v}.
\end{equation}
(For an interpolating field this makes sense even when $m$ is large but
not infinite.) Of course the expansion of eq. (3) is invalid in loops,
where $k$ need not be small, and all four spinor degrees of freedom are
important. Let us denote the remaining degrees of freedom by $\psi_{-v}$, 
\begin{equation}
\slashchar{v} \psi_{-v} = - \psi_{-v}.
\end{equation}
We can refer to $\psi_{+v}$ and $\psi_{-v}$ as the ``particle'' and
``anti-particle'' fields respectively, since eqs. (4,5) are just the
Dirac equations for positive-energy particles and anti-particles with
fixed momentum $mv$. 

We can understand the status of $\psi_{-v}$ by observing that for $k_{\mu}
\ll m$ in (1), 
\begin{equation}
\frac{1 - \slashchar{v}}{2} ~\frac{1}{\slashchar{p} - m + i \epsilon}~
 \frac{1 - \slashchar{v}}{2} =  \frac{1 - \slashchar{v}}{2} ~\frac{1}{2m}
+ {\cal O}(\frac{1}{m^2}).
\end{equation}
We see that relative to the ground state of the heavy quark sector,
 $\psi_{-v}$ is a massive
mode, with mass gap $2m$. Like any heavy mode in field theory we can
integrate out its virtual effects and omit the field from the effective
theory. This is what is done in standard 
HQET, so ${\cal L}_{HQET}$ depends only
on $\psi_{+v}$ and the light quarks and gluons. Normally in field theory
heavy and light particles transform independently under
symmetries. Integrating out the heavy particles does not alter the
symmetry properties of the light particles. RPI is different,
 in that under eq. (2),
$\psi_{\pm v}$ clearly mix. Integrating out the heavy mode $\psi_{-v}$
necessarily complicates RPI in the effective theory.

 In this paper, I derive RPI from first principles, in two versions of
 HQET.
 In section 2, I describe a  version of HQET in which the
anti-particle field $\psi_{-v}$ is explicitly present, but as a non-propagating
auxiliary field.\footnote{ $\psi_{-v}$ plays a role in simplifying RPI
  analogous to the role of auxiliary fields in supersymmetric field
  theory.}
 As in standard HQET, the $1/m$ expansion is performed
{\it before} computing diagrams. The fact that both $\psi_{\pm v}$ are
present makes RPI as simple as it is in the case of heavy spin-$0$
particles. One can in fact write the most general form of the
RPI-constrained effective lagrangian. This is done in section 3. 
In the second step of the program, described in section 4,  the $\psi_{-v}$
field is completely integrated out, resulting in the general form of the
RPI-constrained effective lagrangian in the standard HQET formulation. 
In section 5, I check some of the consequences of RPI on ${\cal
L}_{HQET}$ at the leading orders of $1/m$. Section 6 contains some
discussion and comparison with earlier proposals.   

For simplicity, in this paper I only consider HQET applied to the case
of a single heavy quark in interaction with light quarks and gluons. 
While the character of the arguments and results presented in this paper are
non-perturbative in nature, the proofs are given within perturbation
theory, in the context of minimal subtraction and 
dimensional regularization.  I will address the topic of RPI in
non-perturbative lattice HQET elsewhere.

\section{HQET with an Auxiliary Field}

To focus on the residual momenta $k_{\mu}$ as in eq. (1), we make
the standard field redefinition,
\begin{equation}
\psi_v(x) \equiv e^{i m v.x} \psi(x),
\end{equation}
where $\psi$ is the full QCD quark field and $\psi_v$ will be our HQET
field, and $m$ is the heavy quark mass.  We can
further decompose $\psi_v$,
\begin{equation}
\psi_{\pm v} \equiv \frac{1 \pm \slashchar{v}}{2} \psi_v.
\end{equation}
There is an ambiguity
in just what is meant by the quark ``mass''. The canonical choice is to
choose $m$ so that the {\it residual} mass term for $\psi_{+v}$
vanishes in the HQET.  In general though, any choice is
allowed if it makes the residual mass $\ll m$ (see ref. \cite{falk}).
 The minimal subtraction 
 mass parameter of full QCD will be separately denoted by 
 $\tilde{m}(\mu)$. 

At tree-level, expanding in $1/m$ before or after computing Feynman
diagrams makes no difference, so matching the HQET to full QCD only
amounts to accounting for eq. (7),
\begin{eqnarray}
 {\cal L}^{tree}_{HQET} &=& \overline{\psi}_v e^{i m v.x} (i \slashchar{D} -
\tilde{m}) e^{-imv.x} \psi_v \nonumber \\
&=&  \overline{\psi}_{+v} (iD.v + m - \tilde{m}) \psi_{+v} \nonumber \\
&~& - ~ \overline{\psi}_{-v} (m + \tilde{m} + iD.v) \psi_{-v} \nonumber \\
&~& + ~ \overline{\psi}_{+v} i \slashchar{D}_{\perp} \psi_{-v} 
+ ~  \overline{\psi}_{-v} i \slashchar{D}_{\perp} \psi_{+v},
\end{eqnarray}
where $\slashchar{D}_{\perp} \equiv \slashchar{D} - D.v \slashchar{v}$.
The gauge field and light quark  terms have been suppressed. The
canonical choice for $m$ at this order is clearly $m = \tilde{m}$. 
In using this lagrangian the HQET approach tells us to expand in $1/m ~ 
(1/\tilde{m})$ 
before computing diagrams, so we see that the $\psi_{-v}$
propagator in a background gauge field is really completely local,
\begin{equation}
-\frac{1}{m + \tilde{m} + iD.v} \equiv - \frac{1}{m + \tilde{m}} 
+ \frac{iD.v}{(m + \tilde{m})^2} - ~...
\end{equation}
Order by order in $1/m$ this auxiliary field $\psi_{-v}$ can be
integrated out to yield interactions for the propagating $\psi_{+v}$
field. 

At the quantum level, matching corrections to ${\cal L}_{HQET}$ are
non-trivial because in full QCD we expand in $1/m$ after computing
diagrams while in the HQET we expand the Feynman vertices and
propagators in $1/m$ before computing diagrams. The two steps do not
commute inside loops because of ultraviolet divergences, but since such
divergences are local, order by order in the loop expansion we can
compensate by adding purely local terms to the HQET lagrangian.
To state the matching conditions precisely, we add sources
to full QCD and to the HQET,
\begin{eqnarray}
\delta {\cal L}_{QCD} &=& \overline{\eta} e^{imv.x} \psi + {\rm h.c.}
\nonumber \\
\delta {\cal L}_{HQET} &=& \overline{\eta} \psi_v + {\rm h.c.},
\end{eqnarray}
and demand that the corresponding Green functions agree order
by order in $1/m$ and $\alpha_s$.\footnote{These could be either
  connected or 1PI Green functions. Note that normally the relation
between 1PI vertices in full and effective theories is complicated when
a massive mode is completely integrated out. 
The reason is that a diagram in the full
theory that can be cut in two by cutting an internal massive mode
propagator can become 1PI in the effective theory where the massive mode
is absent. In the present case, as
discussed in the introduction, the massive mode is $\psi_{-v}$, but
because we have left it in the effective theory as an auxiliary field,
we have a simple equality of the 1PI effective actions of the full and
effective theories after matching.}

The most general form for the HQET
lagrangian is,
\begin{eqnarray}
{\cal L}_{HQET} &=& \overline{\psi}_v [S +  \slashchar{V} + 
\sum_{n \geq  2} T^{(n)}.\sigma^{(n)}] \psi_v 
\\
&=& \overline{\psi}_{+v}[S + v.V + 
\sum_{n \geq  2} T^{(n)}.\sigma^{(n)}] \psi_{+v} \nonumber \\
&~& + ~ \overline{\psi}_{-v}[S - v.V  + 
\sum_{n \geq  2} T^{(n)}.\sigma^{(n)}] \psi_{-v} \nonumber \\
&~& + ~ \overline{\psi}_{+v}[\slashchar{V}_{\perp}
 + \sum_{n \geq  2} T^{(n)}.\sigma^{(n)}] \psi_{-v} \nonumber \\
&~& + ~ \overline{\psi}_{-v}[\slashchar{V}_{\perp}  + 
\sum_{n \geq  2} T^{(n)}.\sigma^{(n)}] \psi_{+v},
\end{eqnarray}
where $S, V_{\mu},  T^{(n)}_{\mu_1 ... \mu_n}$ are local hermitian
operators constructed from the light fields, covariant derivatives, and
$v$, and where $\sigma^{(n)}$ is the totally antisymmetrized product of
$n$ $(4 + \epsilon)$-dimensional $\gamma$-matrices,
\begin{eqnarray}
\sigma^{(n)}_{\mu_1 ... \mu_n} &\equiv& \frac{1}{(n!)} \gamma_{[\mu_1} 
\gamma_{\mu_2} ... 
\gamma_{\mu_n ]}; ~~ n = 4k-1, 4k-2, \nonumber  \\
\sigma^{(n)}_{\mu_1 ... \mu_n} &\equiv& \frac{i}{(n!)} \gamma_{[\mu_1} 
\gamma_{\mu_2} ... 
\gamma_{\mu_n ]}; ~~ n = 4k, 4k-3.
\end{eqnarray}
In four dimensions  $\sigma^{(3)}$ and  $\sigma^{(4)}$ can be
reduced by introducing   $\gamma_5$, but I have refrained from doing this
to avoid problems of  dimensionally continuing $\gamma_5$. In four
dimensions we also have $\sigma^{(n)} = 0$ for $n \geq 5$, so they
correspond to {\it evanescent} operators, only occurring in the
counterterm lagrangian of the effective theory.\footnote{The reason the
  evanescent operators {\it only} appear among  the minimal subtraction
  counterterms is just the generalization of the arguments of
  ref. \cite{evanesce}. It is possible that the technical 
problem of evanescent  operators can be avoided by using {\it
  dimensional reduction} as the regulator. See for example the discussion in
ref. \cite{gimenez}. I have not investigated this possibility in the
present context.}

The absence of operators higher than bilinear in
the heavy quark field is because such vertices make no contribution to
(dimensionally regulated) amplitudes in the effective theory describing
a single heavy quark. 

The different Dirac structures are of different orders in $1/m$, which
can be worked out by finding the lowest dimension 
operators that can occur in HQET with these structures,
 and respecting QCD symmetries. However one of these symmetries is RPI
 itself, which is discussed next. This power-counting
 exercise is therefore deferred till the end of the next section.

\section{Reparameterization Invariance}

The HQET with the auxiliary anti-particle field satisfies a very simple
form of RPI, ${\cal L}_{HQET}$ being invariant under 
\begin{equation}
v \rightarrow v + \delta v; ~~ \psi_{v}(x) \rightarrow e^{i m \delta
  v.x} \psi_{v}(x). 
\end{equation}
At tree-level for example, this invariance is manifest in the first line
of eq. (9).   
Beyond tree-level the proof of RPI goes as follows.
 By eq. (7), the quark field of full QCD, 
$\psi(x)$, is invariant under eq. (15). This means ${\cal L}_{QCD}$ is
invariant except for the source term, eq. (11). Clearly the QCD
partition functional then satisfies, 
\begin{equation}
Z[\eta, \overline{\eta}, J_{\mu}, v] = 
Z[e^{i m \delta v.x} \eta, \overline{\eta} e^{- i m \delta v.x},
J_{\mu}, v + \delta v],
\end{equation}
where $J_{\mu}$ is the gauge field source. This relation holds even when
the functional integral is dimensionally regulated and minimal
subtraction counterterms added. (Dimensonal regularization is a ``good''
regulator for reparameterizations, eq. (15).) After matching, this RPI
and renormalized partition functional is equated (to whatever order in
$\alpha_s$ and $1/m$ one is working) to the renormalized HQET partition
functional,
\begin{equation}
Z[\eta, \overline{\eta}, J_{\mu}, v] = \int {\cal D} A_{\mu} {\cal
  D} \psi_v {\cal D} \overline{\psi}_v e^{i\int d^{4 + \epsilon}x
~ [{\cal L}_{HQET} + {\cal L}_{c.t.}  + J.A + \overline{\eta} \psi_v +
\overline{\psi}_v \eta]},
\end{equation}
where ${\cal L}_{c.t.}$ denotes the HQET minimal subtraction
counterterms, and the Fadeev-Popov ghost determinant is subsumed into
the gauge field measure. Noting that the source term, eq. (11), and the
regulated measure are both invariant under $\eta \rightarrow 
e^{i m \delta v.x} \eta, \psi_v \rightarrow e^{i m \delta v.x} \psi_v$,
we see that
\begin{eqnarray}
 Z[e^{i m \delta v.x} \eta, \overline{\eta} e^{-i m \delta v.x}, J_{\mu},
v + \delta v] = ~~~~~~~~~~~~~~~~~~~~~~~~~~~~~~~~~~~~~~~~~~~~~~~~ \nonumber \\
\int {\cal D} A_{\mu} {\cal
  D} \psi_v {\cal D} \overline{\psi}_v e^{i\int d^{4 + \epsilon}x
~ [{\cal L}_{HQET} + \delta {\cal L}_{HQET} + {\cal L}_{c.t.} + \delta 
{\cal L}_{c.t.}  + J.A + \overline{\eta} \psi_v +
\overline{\psi}_v \eta]},
\end{eqnarray}
where,
\begin{equation}
\delta {\cal L} \equiv 
{\cal L}(e^{i m \delta v.x} \psi_v, \overline{\psi}_v e^{-i m \delta
  v.x}, A_{\mu}, v + \delta v) - {\cal L}(\psi_v, 
\overline{\psi}_v, A_{\mu}, v).
\end{equation}
Eq. (16) therefore reads,
\begin{equation}
\int {\cal D} A_{\mu} {\cal
  D} \psi_v {\cal D} \overline{\psi}_v ~ \int dx
(\delta {\cal L}_{HQET} + \delta 
{\cal L}_{c.t.}) e^{i \int dx ~ {\cal L}_{HQET} + {\cal L}_{c.t.}  + J.A +
 \overline{\eta} \psi_v + \overline{\psi}_v \eta} = 0.
\end{equation}
It follows that 
\begin{equation}
\int dx(\delta {\cal L}_{HQET} + \delta {\cal L}_{c.t.}) = 0,
\end{equation}
since all insertions of this operator into Green functions vanish. In
minimal subtraction we separately have, $\delta {\cal L}_{HQET} = 0$ and 
$ \delta {\cal L}_{c.t.} = 0$ since they have distinct
$1/\epsilon$-dependence. This proves the RPI of ${\cal L}_{HQET}$.

The RPI constraint on ${\cal L}_{HQET}$ has a simple solution: $v_{\mu}$
and $D_{\mu}$ can only occur in the lagrangian with $\overline{\psi}_v,
\psi_v$ (see eq. (12)), in the combination $v_{\mu} + i
D_{\mu}/m$. (This is only true for covariant derivatives acting on the
heavy fields, not on the light fields, which are trivially RPI.) 
That is, $S, V_{\mu}, T^{(n)}$ are
constructed out of $v_{\mu} + i D_{\mu}/m$.
The proof is straightforward. Instead of working with  $\overline{\psi}_v,
\psi_v$ directly, we can write any effective lagrangian 
 in terms of the combinations
$\overline{\psi}_v e^{i m v.x}$ and $e^{-i m v.x} \psi_v$ which, as we
noted earlier,  are invariants under eq. (15). A general term of 
${\cal L}_{HQET}$ has the
form $\overline{\psi}_v e^{i m v.x} f(v_{\mu}, D_{\nu}) e^{-i m v.x}
\psi_v$ where $f$ is some monomial in its arguments (and can be a
matrix in spinor, Lorentz or color spaces), multiplied by a term involving 
only light fields and their covariant derivatives. In this decomposition
{\it only} the  $v_{\mu}$ argument of $f$ transforms under
eq. (15). Furthermore the only invariant combination of $v_{\mu}$'s alone is
trivial, $v^2 = 1$. Therefore imposing RPI just translates into the
$v$-independence of $f$. The phase factors can be cancelled against each
other by noting that $D_{\mu} 
 e^{i m  v.x} ... =  e^{i m  v.x} (i m v_{\mu} + D_{\mu}) ...$ . 
Thus, the general RPI term in   ${\cal  L}_{HQET}$ has the form 
$\overline{\psi}_v f(i m v_{\mu} +  D_{\mu}) \psi_v$, multiplied by a
factor consisting of light fields, thereby completing the proof. 
We see that in the auxiliary
field formulation, RPI is as simple as in theories with 
spin-$0$ heavy quarks \cite{luke}. As emphasized by Luke and Manohar,
the central reason for the complication of RPI in the standard
formulation of HQET is that  the heavy quark field must satisfy a
$v$-dependent constraint. By including the auxiliary field in the
present formulation, there is no constraint on $\psi_v = \psi_{+v} +
\psi_{-v}$.

We are now in a position to give the power-counting for the operators of
the various Dirac structures, described at the end of the last
section. They are restricted by conventional QCD symmetries, and by
RPI. The $T^{(n)}$ must have at least one pair of anti-symmetrized
Lorentz indices to be contracted with the $\sigma^{(n)}$. These indices
can only arise from a pair of reparameterization-covariant derivatives,
$v_{\mu} + i D_{\mu}/m$. That is, 
$T^{(n)} = ... (v + iD/m)_{[\mu} ... (v + iD/m)_{\nu]} ...~$. However,
terms in which any $v$'s are pulled out of the parentheses do not
survive the anti-symmetrization unless there are also 
derivatives between the
two sets of parentheses. It follows that the $T^{(n)}$ operators must
contain at least two derivatives, $D_{\mu}$. So by dimensional
analysis the $T^{(n)}$ can be at most ${\cal O}(1/m)$. The lowest
dimension operator that can appear in $S$ is the unit operator, with a
coefficient of order $m$ by dimensional analysis. Similarly the lowest
dimension operator that can appear in $V_{\mu}$ is $v_{\mu} + i
D_{\mu}/m$, again with a coefficient of order $m$.  
 There is just one   subtlety which  can already be seen at
tree-level. Though $S$ and $V_{\mu}$ are both of order $m$, 
$S + v.V \sim {\cal O}(1)$ if we choose the residual
mass to vanish (or be at most ${\cal O}(\Lambda_{QCD})$). To summarize,
\begin{eqnarray}
S - v.V \sim {\cal O}(m); ~~ S + v.V \sim {\cal O}(1);
 ~~ \slashchar{V}_{\perp} \sim {\cal O}(1); ~~ T^{(n)} \leq {\cal O}(1/m). 
\end{eqnarray}

\section{Integrating out the Anti-Particle Field}

The auxiliary field formulation of HQET is a valid
calculational scheme in its own right, once matched to full
QCD. However, having solved the RPI constraint in this formulation we
are free to integrate out the auxiliary field and
 return to the
standard formulation of HQET in terms of $\psi_{+v}$ (once we turn off
the source for $\psi_{-v}$). The result will be
the general RPI-constrained form of the standard HQET lagrangian. Since
$\psi_{-v}$ occurs at most quadratically in eq. (13) we can easily
integrate it out to get,
\begin{eqnarray}
&~&  {\cal L}_{HQET} + {\cal L}_{c.t.} = \overline{\psi}_{+v} \{S + v.V +
\sum_{n \geq  2} T^{(n)}.\sigma^{(n)}
- [\slashchar{V}_{\perp} + \sum_{n \geq  2} T^{(n)}.\sigma^{(n)}] \nonumber \\
&\times&
[S - v.V + \sum_{n \geq  2} T^{(n)}.\sigma^{(n)}]_{(-v)}^{-1}
 [\slashchar{V}_{\perp} + \sum_{n \geq  2} T^{(n)}.\sigma^{(n)}]\}\psi_{+v}.
\end{eqnarray}
This expression requires some explanation. The $(-v)$ subscript
appearing in the operator inverse is an instruction to perform the
inversion completely within the $(1 - \slashchar{v})/2$ subspace, not the
full spinor space, thereby getting a result that also lives in the
subspace.  Secondly, this inverse is really
local. The reason is that the dominant part of $S - v.V$ is a constant
of order $m$. Therefore expanding the operator inverse in powers of
$1/m$ yields a series of local operators, similarly to eq. (10). The
last observation concerns renormalization. When integrating $\psi_{-v}$
out of the auxiliary field formulation,
 the effective lagrangian in the functional integral contains
counterterms. Therefore each operator  ${\cal O}_0 = S, 
V, T^{(n)}$ appearing in eq. (23)  has the form,
\begin{equation}
{\cal O}_0 =  {\cal O}(\mu) + {\cal O}_{poles},
\end{equation}
where ${\cal O}_{poles}$ is a sum of $1/\epsilon^n$ pole counterterms
obtained in the auxiliary field formulation of HQET. The standard HQET
lagrangian is just the finite part of eq. (23), the poles providing the
counterterms for the standard HQET formulation. Clearly, 
given eq. (24), to all orders in $\alpha_s$ and $1/m$ this finite part must
also have the form of eq. (24), where the operators are the ${\cal
  O}(\mu)$.\footnote{The products of $(4 + \epsilon)$-dimensional
$\gamma$-matrices encountered in eq. (23) can be re-expressed as linear
combinations of the fully anti-symmetrized products, $1, \gamma_{\mu},
\sigma^{(n)}$, and simplified using $(1 + \slashchar{v})/2$-projected
identities to arrive at a canonical form for the standard HQET
lagrangian. These particular manipulations do not introduce any explicit
powers of $\epsilon$, which would have changed the separation of the
effective lagrangian into the finite part and minimal subtraction
counterterms.}
 That is, both the renormalized and unrenormalized standard HQET
lagrangians have the form of eq. (23).  
RPI restricts the ${\cal O}(\mu)$ and the ${\cal O}_{poles}$ to be
constructed from $v_{\mu} + iD_{\mu}/m$. 
This result agrees with ref. \cite{kilian}.

We can view the form of
eq. (23) as the solution to demanding invariance 
under a reparameterization transformation of $\psi_{+v}$. We can derive
this transformation by noting that the Gaussian integral performed to
eliminate $\psi_{-v}$ is the same as using the $\psi_{-v}$ ``equation of
motion'' following from eq. (13), 
\begin{equation}
\psi_{-v} = -~ [S - v.V + 
\sum_{n \geq  2} T^{(n)}.\sigma^{(n)}]_{(-v)}^{-1}
 [\slashchar{V}_{\perp} + \sum_{n \geq  2} T^{(n)}.\sigma^{(n)}] \psi_{+v}.
\end{equation}
(See the remarks after eq. (23) regarding the interpretation of the
operator inverse). Thus the $\psi_v$ reparameterization transformation in
the  auxiliary field formulation corresponds to the following
transformation in standard HQET, 
\begin{eqnarray}
\psi_{+v'} &=& \frac{1 + \slashchar{v}'}{2} e^{i m \delta v.x} \psi_{v}
\nonumber \\
&=& e^{i m \delta v.x} \frac{1 + \slashchar{v}'}{2} \{1 - \nonumber \\
&~& 
[S - v.V + \sum_{n \geq  2} T^{(n)}.\sigma^{(n)}]_{(-v)}^{-1}
 [\slashchar{V}_{\perp} + \sum_{n \geq  2} T^{(n)}.\sigma^{(n)}]\}\psi_{+v}
 \nonumber \\
&=& \{1 + i m \delta v.x + \frac{\delta \slashchar{v}}{2} 
- \frac{
\delta \slashchar{v}}{2} \times \nonumber \\
&~&
[S - v.V + \sum_{n \geq  2} T^{(n)}.\sigma^{(n)}]_{(-v)}^{-1}
 [\slashchar{V}_{\perp} + \sum_{n \geq  2} T^{(n)}.\sigma^{(n)}]\}\psi_{+v},
\end{eqnarray}
where ${\cal O}(\delta v^2)$ terms have been dropped in the last
line. This general form also agrees with Kilian and Ohl \cite{kilian}.
However,  the
{\it specific} forms of $S, V,  T^{(n)}$  must be
determined by matching the auxiliary field formulation  of HQET to full
QCD. It is straightforward to see that the operators appearing in
eq. (26) are either the ${\cal O}(\mu)$ or ${\cal O}_0$ depending on
whether one is considering RPI of ${\cal L}_{HQET}$ or 
${\cal L}_{HQET} + {\cal L}_{c.t.}$ respectively.

\section{Examples in Standard HQET}

It is a well-believed expectation that RPI should constrain the term in
the standard 
HQET lagrangian, $-\frac{1}{2m} \overline{\psi}_{+v} D^2 \psi_{+v}$,
to have unit coefficient to all orders in $\alpha_s$ (for example the
analogous statement
is straightforward to prove for heavy scalars \cite{luke}).
 Let us see how this
emerges from the general form, eq. (23). One can easily
 check that the $T^{(n)}$  can
have no effect on the ${\cal O}(1/m)$ operator of interest. Thus the
part of the lagrangian with the right Dirac structure to this order is
given by 
\begin{eqnarray}
{\cal L}_{HQET} &=& \overline{\psi}_{+v} \{S + v.V 
- \slashchar{V}_{\perp}(S - v.V)^{-1} \slashchar{V}_{\perp} \}\psi_{+v}
+ ... ~ ,
\end{eqnarray}
where $S$ is a  function $f(v + iD/m)^2)) = f(1 + 2iv.D/m -
D^2/m^2)$, and $V_{\mu} = g(1 + 2iv.D/m - D^2/m^2)(v_{\mu} + i
D_{\mu}/m)$. Actually, RPI permits more general forms for $S$ and $V$
which can be reduced to the above by using the fact that $v_{\mu} +
iD_{\mu}/m$ commutes with $v_{\nu} + iD_{\nu}/m$ up to terms involving
the gauge field strength, such terms being discarded into the ellipsis
above. To order $1/m$ the general forms of $S$ and $V_{\mu}$ are, 
\begin{eqnarray}
S &=& -m_1 + Z_1 (iv.D - \frac{D^2}{2m}) + k_1 \frac{(v.D)^2}{m}
\nonumber \\
 V_{\mu} &=&
[m_2 + Z_2 (iv.D - \frac{D^2}{2m}) + k_2 \frac{(v.D)^2}{m}](v_{\mu} + i
\frac{D_{\mu}}{m}). 
\end{eqnarray}
Substituting into eq. (27),
\begin{eqnarray}
&~& {\cal L}_{HQET} = \overline{\psi}_{+v}\{m_2 - m_1 \nonumber \\
&+& (Z_1 + Z_2)(iv.D - 
\frac{D^2}{2m}) + \frac{m_2}{m} iv.D - \frac{m_2^2 D^2}{m^2 (m_1 + m_2)}
\}\psi_{+v} + ... 
\end{eqnarray}
It remains to put ${\cal L}_{HQET}$ into canonical form. The $m_2 - m_1$
term is the ``residual mass'' term. We  choose $m$ to make the
residual mass vanish, $m_1 = m_2$. Next we perform  wavefunction
renormalization to ensure that $\overline{\psi}_{+v} iv.D \psi_{+v}$ has
canonical unit coefficient. The  result is 
\begin{equation}
{\cal L}_{HQET} = \overline{\psi}_{+v}(iv.D - 
\frac{D^2}{2m})\psi_{+v} + ... ~, 
\end{equation}
as we wished to prove.

A second example involves  spin-dependent terms in the standard
effective lagrangian. Both Chen's and Luke and Manohar's versions of RPI
predict that the coefficients of the operators,
\begin{eqnarray}
O_{mag} &=& \frac{g_s}{4m} \overline{\psi}_{+v} \sigma^{\mu \nu} G_{\mu
  \nu} \psi_{+v}, \nonumber \\
O_2 &=&  \frac{i g_s}{8m^2} \overline{\psi}_{+v} \sigma^{\alpha \mu}
v^{\nu} \{D_{\alpha}, G_{\mu \nu} \} \psi_{+v},
\end{eqnarray}
are related by 
\begin{equation}
2C_{mag} = C_2 + 1
\end{equation}
(when the heavy quark kinetic term and residual mass terms are put into
canonical form) \cite{luke} \cite{chen}
\cite{finkemeier}. Ref. \cite{finkemeier} showed this to be true
independently of the earlier RPI proposals. 
 It is therefore of interest to know whether eq. (32) follows
from our general form, eq. (23). The answer is yes, as a
 straightforward but tedious calculation 
(along similar lines to the last example) shows.

\section{Discussion}

At tree-level, $S = - \tilde{m}, V_{\mu} = m v_{\mu} + i
D_{\mu}, T^{(n)} = 0$, and if we choose $m = \tilde{m}$, eq. (26)
becomes the transformation proposed by Chen \cite{chen}. Beyond
tree-level, there are definitely matching corrections to $S, V, T^{(n)}$ in the
auxiliary field formulation of HQET, so the reparameterization
transformation of standard HQET, eq. (26), is necessarily corrected 
beyond Chen's tree-level form. This means that Chen's proposed form of
the RPI transformation must be wrong. Therefore at first sight, it is
surprising that  Finkemeier,
Georgi and McIrvin \cite{finkemeier} 
could prove all the predictions of  Chen's form of RPI to ${\cal
  O}(1/m^2)$ in ${\cal L}_{HQET}$, such as eq. (32) for example. We are
in a position to understand this. The results to this order in
the effective lagrangian are completely determined by the form of the
reparameterization transformation to order $1/m$, so that the $T^{(n)}$ 
are irrelevant inside  eq. (26), by eq. (22), 
and we can use the forms in eq. (28) for
$S$ and $V$. When we enforce the vanishing of the residual mass, $m_1 =
m_2$, we find,
\begin{eqnarray}
\psi_{+v'}
= \{1 + i m \delta v.x + \frac{\delta \slashchar{v}}{2} 
+ \frac{
\delta \slashchar{v}}{2} ~ \frac{i \slashchar{D}_{\perp}}{2m}
\}\psi_{+v} + {\cal O}(1/m^2). 
\end{eqnarray}
This is precisely the same as Chen's transformation to order $1/m$,
thereby explaining the success of Chen's proposal to order $1/m^2$ in
the effective lagrangian. Beyond this order, eq. (26) and Chen's
transformation no longer coincide, and only eq. (26) is correct.

Eq. (26) is also not the  RPI transformation proposed by Luke and
Manohar \cite{luke}. However to order $1/m^2$, 
 ref. \cite{finkemeier} has shown that there
is a field redefinition (which does not affect the S-matrix) which, 
when  compounded with the Luke-Manohar
transformation, yields eq. (33). It is possible this agreement
 with eq. (26) up to
field redefinitions persists at higher orders in
HQET. See ref. \cite{manohar} for some more discussion.

The auxiliary field formulation of HQET is a viable calculational scheme
which makes RPI extremely simple, given by eq. (15). 
If one prefers the standard
formulation of HQET, the correct form of RPI is the one 
 obtained by Kilian and Ohl
\cite{kilian} and rederived in this paper.

\section*{Acknowledgments}
This research was supported by the U.S. Department of Energy under grant
\#DE-FG02-94ER40818.
I am grateful for useful discussions with Markus Finkemeier, Howard
Georgi, Markus Luty,  and Matt McIrvin.


\begin{thebibliography}{99}
\bibitem{grinstein} B. Grinstein, Annu. Rev. Nucl. Part. Sci. 42 (1992)
  101.
\bibitem{dugan} M. J. Dugan, M. Golden and B. Grinstein,
  Phys. Lett. B282 (1992) 142. 
\bibitem{luke} M. Luke and A. V. Manohar, Phys. Lett. B286 (1992) 348. 
\bibitem{chen} Y.-Q Chen, Phys. Lett. B317 (1993) 421.
\bibitem{manohar} A. Manohar, preprint UCSD/PTH 97-01, hep-ph/9701294.
\bibitem{kilian} W. Kilian and T. Ohl, Phys. Rev. D50 (1994) 4649. 
\bibitem{finkemeier} M. Finkemeier, H. Georgi and M. McIrvin, preprint
  HUTP-96/A053, hep-ph/9701243.
\bibitem{falk} A. Falk, M. Luke and M. Neubert, Nucl. Phys. B388 (1992) 363.
\bibitem{evanesce} M. J. Dugan and B. Grinstein, Phys. Lett. B256 (1991)
  239.
\bibitem{gimenez} V. Gimenez, Nucl. Phys. B401 (1993) 116.
\end{thebibliography}
\end{document}